\documentclass[english]{nature}
\usepackage[T1]{fontenc}
\usepackage[latin9]{inputenc}
\usepackage{geometry}
\usepackage{amsmath}
\usepackage{amssymb}
\usepackage{graphicx}
\usepackage{subscript}

\makeatletter
\pdfoutput=1

\makeatother

\usepackage{babel}
\begin{document}

\title{Electron--hole correlations govern Auger recombination in nanostructures }

\author{John P. Philbin$^{*,1}$ \& Eran Rabani$^{*,1,2,3}$}
\maketitle
\begin{affiliations}
\item Department of Chemistry, University of California, Berkeley, California
94720, United States
\item Materials Science Division, Lawrence Berkeley National Laboratory,
Berkeley, California 94720, United States
\item The Sackler Center for Computational Molecular and Materials Science,
Tel Aviv University, Tel Aviv, Israel 69978
\end{affiliations}
\begin{abstract}
The fast nonradiative decay of multiexcitonic states via Auger recombination
is a fundamental process affecting a variety of applications based
on semiconductor nanostructures. From a theoretical perspective, the
description of Auger recombination in confined semiconductor nanostructures
is a challenging task due to the large number of valance electrons
and exponentially growing number of excited excitonic and biexcitonic
states that are coupled by the Coulomb interaction. These challenges
have restricted the treatment of Auger recombination to simple, noninteracting
electron--hole models. Herein we present a novel approach for calculating
Auger recombination lifetimes in confined nanostructures having thousands
to tens of thousands of electrons, explicitly including electron--hole
interactions. We demonstrate that the inclusion of electron--hole
correlations are imperative to capture the correct scaling of the
Auger recombination lifetime with the size and shape of the nanostructure.
In addition, correlation effects are required to obtain quantitatively
accurate lifetimes even for systems smaller than the exciton Bohr
radius. Neglecting such correlations can result in lifetimes that
are $2$ orders of magnitude too long. We establish the utility of
the new approach for CdSe quantum dots of varying sizes and for CdSe
nanorods of varying diameters and lengths. Our new approach is the
first theoretical method to postdict the experimentally known ``universal
volume scaling law'' for quantum dots and makes novel predictions
for the scaling of the Auger recombination lifetimes in nanorods. 
\end{abstract}
\begin{figure}[t]
\begin{centering}
\includegraphics{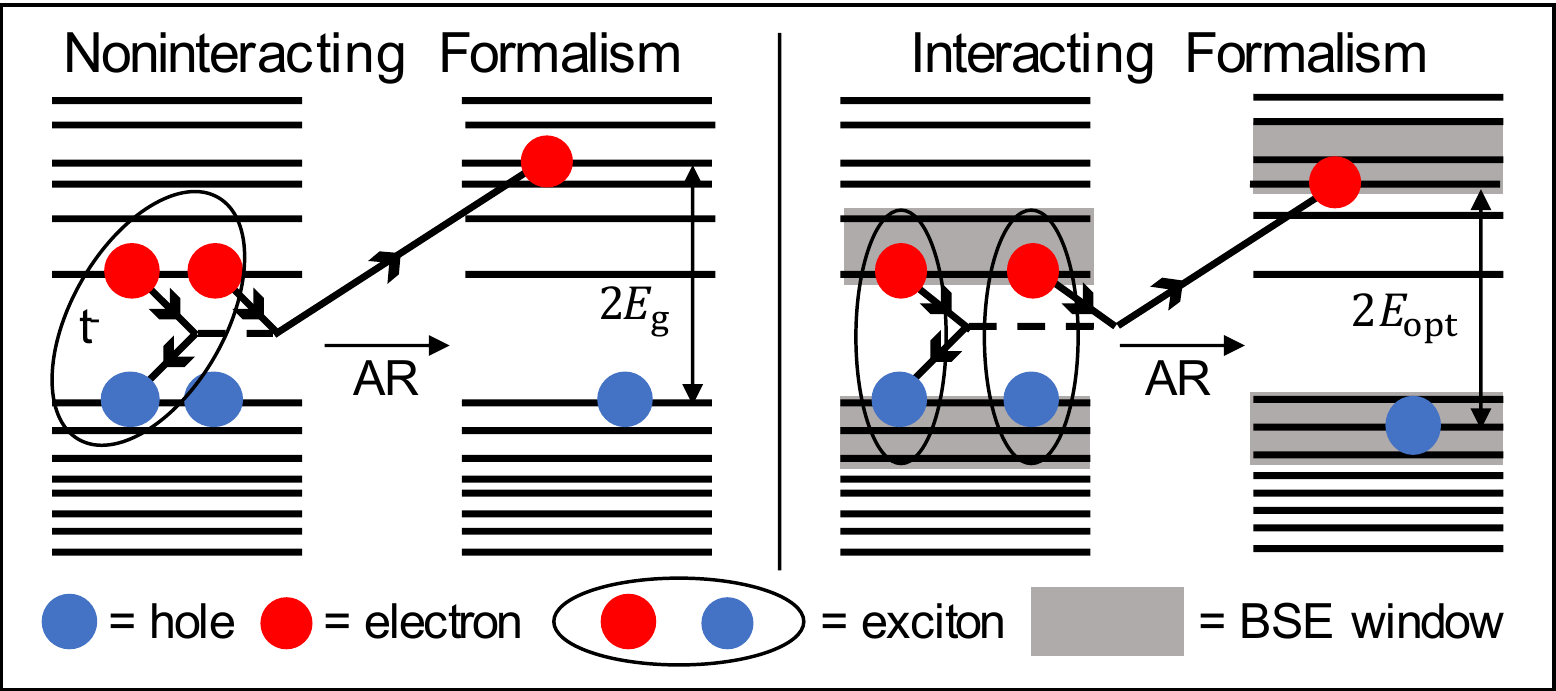}
\par\end{centering}
\caption{\label{fig:ar-picture}Pictorial representations are shown for the
electron channel of an Auger recombination (AR) event in the noninteracting
(left) and interacting (right) formalisms. The black horizontal lines
represent the discrete quasiparticle states of the semiconductor nanostructures.
The gray box in the interacting formalism represents the fact that
the excitons (correlated electron--hole pairs) are a linear combination
of the quasiparticle states within the box that were included in the
BSE. $E_{\text{g}}$ is the fundamental gap and $E_{\text{opt}}$
is the optical gap. $\left|B\right\rangle ^{\left(0\right)}$ is the
initial state in the noninteracting formalism (note that one of the
holes is a spectator and the AR process describes a negative trion,
$t^{-}$, decaying to an excited quasielectron state). $\left|B\right\rangle $
is the initial state in the interacting formalism composed of two
excitons and all $4$ particles are involved in the AR process. The
final states in both formalisms are given by $\left|S\right\rangle ^{\left(0\right)}$.
The dashed line represents the Coulomb interaction.}

\end{figure}

The fast nonradiative decay of multiexcitonic states is a central
process to many nanocrystal--based applications.\cite{Klimov2000,Klimov2014}
This nonradiative decay occurs primarily via Auger recombination (AR)
in which one electron--hole pair recombines by transferring its energy
to an additional charge carrier (Fig.~\eqref{fig:ar-picture}). In
some cases, such as light harvesting devices, AR can limit  performance
by rapidly quenching the photoluminescence~\cite{Colvin1994,Klimov2000,Imamoglu2000,Pietryga2008,Bae2013}
and destroying the population inversion required for nanocrystal based
lasers,\cite{Fathpour2004} while in other cases, such as photodetectors,\cite{Sukhovatkin2009}
single photon sources~\cite{Nair2011} and even for photocatalysis,\cite{Ben-Shahar2018}
it can improve performance by providing a source of hot electrons.
Therefore, developing a unified framework to describe AR is important
from both fundamental and applied perspectives. 

In recent years, much effort has been put into -- and much success
obtained in -- the development of synthetic techniques and principles
that result in nanocrystals (NCs) with rationally designed AR lifetimes.\cite{Klimov2014}
Synthesizing giant NCs offers the simplest and most well--known approach
to increase the AR lifetime. This approach works well because the
AR lifetime, $\tau_{\text{AR}}$, in single--material quantum dots
(QDs) obeys the ``universal volume scaling law'' (i.e\emph{.}, $\tau_{\text{AR,QD}}\propto V$
in QDs).\cite{Klimov2000,Robel2009,Garcia-Santamaria2009,Pijpers2010}
However, current theories predict a steeper scaling with the QD volume,\cite{Chepic1990,Vaxenburg2015,Vaxenburg2016}
signifying only a partial understanding of the AR process even in
spherical, 0D NCs. In addition to controlling the AR lifetime by changing
the system size, many reports have found that an intelligent design
of core/shell NCs with sharp or gradual interfaces allows for the
AR lifetimes in NCs to be tuned.\cite{Cragg2010,Zavelani-Rossi2010,Garcia-Santamaria2011,Park2014,Jain2016,Vaxenburg2016,Jain2017} 

The situation is somewhat more confusing for non--spherical NCs.\cite{Htoon2003,Taguchi2011,Zhu2012,Yang2012,Padilha2013,Aerts2013,She2015,Stolle2017,Li2017}
The AR lifetime in 1D nanorod (NR) structures was reported to scale
linearly with the length ($L$) of the NRs (i.e., $\tau_{\text{AR,NR}}\propto L$),
but this observation has not been derived from first principles. \textit{\emph{R}}ecently,
it was argued that the AR decay in PbSe NRs has a crossover from cubic
to bimolecular scattering as the length of the NR is increased,\cite{Aerts2013}
calling into question the monotonic length dependence. Further complications
arise from the difficulty to measure precisely the AR lifetimes~\cite{Taguchi2011}
and also to independently control the dimensions of NRs by current
synthetic techniques. In fact, it was shown that NRs of equal volume
(but differing diameters and lengths) can have AR lifetimes that differ
by more than a factor of $2$,\cite{Padilha2013} but whether this
indicates a deviation from the volume scaling observed in QDs remains
an open question. 

Nanoplatelet (NPL) structures appear to provide an example of the
breakdown of the volume scaling of AR lifetimes. Contradictory results
have been reported for the scaling of AR lifetimes with the lateral
area ($A$). She \emph{et al.} showed that the AR lifetimes are independent
of $A$,\cite{She2015} while recently it was argued to scale linearly
with $A$, attributed to collisions of excitons limited by their spatial
diffusion.\cite{Li2017} The scaling of the AR lifetime as a function
of the number of monolayers (ML) was reported to obey a seventh power
dependence, $\tau_{\text{AR,NPL}}\propto\left(\text{ML}\right)^{7},$
in CdSe NPLs.\cite{Li2017} This was rationalized by a simple noninteracting
effective mass model.\cite{Li2017}

In order to simplify and better understand the size and dimensionality
dependence of AR lifetimes in NCs, a unified theoretical framework
for calculating AR lifetimes in 0D, 1D and 2D nanostructures must
be developed. Such a development has been hampered by various factors,
including limitations resulting from the enormous number of excitonic
and biexcitonic states in NCs as well as the difficulties in including
electron--hole correlation effects. Indeed, previous theoretical
works have relied on a non--atomistic model~\cite{Chepic1990,Wang2006}
or a noninteracting electron--hole picture, thought to be suitable
for strongly confined systems.\cite{Chepic1990,Wang2003,Cragg2010,Korkusinski2011,Vaxenburg2015,Vaxenburg2016}
However, this approach fails to handle the continuous transition from
strong to weak confinement regimes as well as nanostructures that
have both strong and weak confinement along different dimensions (e.g.,
weakly confined along the NR axis and strongly confined in the others). 

In this Letter, we develop a unified approach for calculating AR lifetimes
that is applicable to all degrees of confinement. The approach is
based on Fermi's golden rule to couple excitonic with biexcitonic
states. Electron--hole correlations are explicitly included in the
initial biexcitonic states by solving the Bethe--Salpeter equation
(BSE) to obtain correlated electron--hole states which are then used
to form the initial biexcitonic states. This procedure captures most
of the electron--hole correlation as the exciton binding energy is
typically an order of magnitude larger than the biexciton binding
energy.\cite{Patton2003} Through a study of CdSe QDs and NRs of varying
dimensions, we show that our approach predicts AR lifetimes in quantitative
agreement with experiments whereas the noninteracting formalism often
overestimates the AR lifetimes by $1-2$ orders of magnitude. The
shorter AR lifetimes are a consequence of electron--hole pair localization
which increases the Coulomb coupling and thereby the AR rate in the
interacting formalism. By comparing the interacting and noninteracting
formalisms (Fig.~\eqref{fig:ar-picture}), we also make evident the
importance of including electron--hole correlations for the first
theoretical postdiction of the observed volume scaling of the AR lifetime
in QDs. Interestingly, the transition to the regime where excitonic
effects must be included for an accurate AR lifetime calculation occurs
at a surprisingly small diameter in CdSe QDs, below the exciton Bohr
radius of CdSe. Additionally, we explain the AR lifetime scaling behavior
in terms of the scaling of the Coulomb matrix elements and the density
of final states in QDs and NRs. The method presented in this Letter
is generally applicable to 0D, 1D, 2D and NC heterostructures.

AR involves the coupling of an initial biexcitonic state ($\left|B\right\rangle $)
of energy $E_{B}$ to a final excitonic state ($\left|S\right\rangle $)
of energy $E_{S}$ via the Coulomb interaction ($V$). We utilize
Fermi's golden rule to calculate the AR lifetime ($\tau_{\text{AR}}$)
where we average over thermally distributed initial biexcitonic states
and sum over all final decay channels into single excitonic states:
\begin{eqnarray}
\tau_{\text{AR}}^{-1} & = & \sum_{B}\frac{e^{-\beta E_{B}}}{Z_{B}}\left[\frac{2\pi}{\hbar}\sum_{S}\left|\left\langle B\left|V\right|S\right\rangle \right|^{2}\delta\left(E_{B}-E_{S}\right)\right].\label{eq:fermis_golden_rule}
\end{eqnarray}
In the above, the delta function $\delta\left(E_{B}-E_{S}\right)$
enforces energy conservation between the initial and final states
and $Z_{B}$ is the partition function for biexcitonic states. Note
that later when we compare to experimental values, we use a room temperature
$\beta$ for this Boltzmann weighted average, but we do not include
temperature fluctuations in our NC configurations.\cite{Balan2017} 

A brute force application of equation~\eqref{eq:fermis_golden_rule}
for nanostructures is prohibitive for several reasons. First, there
is currently no tractable electronic structure method for a fully--correlated
biexcitonic state and for excitonic states at high energies. Second,
the number of initial and final states that satisfy energy conservation
increases rapidly with the system size. For these reasons, computational
and theoretical studies of AR in confined nanostructures have relied
on a noninteracting formalism to describe $\left|S\right\rangle $
and $\left|B\right\rangle $:\cite{Chepic1990,Wang2003,Wang2006,Cragg2010,Korkusinski2011,Vaxenburg2015,Vaxenburg2016}
\begin{eqnarray}
\left|S\right\rangle ^{(0)} & = & a_{a}^{\dagger}a_{i}\left|0\right\rangle \otimes\left|\chi_{S}\right\rangle \label{eq:S NI}\\
\left|B\right\rangle ^{(0)} & = & a_{b}^{\dagger}a_{j}a_{c}^{\dagger}a_{k}\left|0\right\rangle \otimes\left|\chi_{B}\right\rangle ,\label{eq:B NI}
\end{eqnarray}
where the superscript ``$(0)$'' signifies a noninteracting picture
is used. In the above, $a_{a}^{\dagger}$ and $a_{i}$ are electron
creation and annihilation operators in quasiparticle state ``$a$''
and ``$i$'', respectively. The indexes $a,b,c...$ refer to the
quasiparticle electron (unoccupied) states and $i,j,k...$ refer to
quasiparticle hole (occupied) states, with corresponding quasiparticle
energies $\varepsilon_{a}$ and $\varepsilon_{i}$. In equation~\eqref{eq:B NI},
$\left|0\right\rangle $ is the ground state and $\left|\chi_{S}\right\rangle $
and $\left|\chi_{B}\right\rangle $ are the spin parts of the wavefunctions
for excitons and biexcitons, respectively. Within the noninteracting
formalism, the excitonic and biexcitonic energies are given by $E_{S}^{(0)}=\varepsilon_{a}-\varepsilon_{i}$
and $E_{B}^{(0)}=\varepsilon_{b}-\varepsilon_{j}+\varepsilon_{c}-\varepsilon_{k}$,
respectively. The AR lifetime takes an explicit form (see the Methods
section for a detailed derivation and discussion of the spin states
studied herein) given by:
\begin{eqnarray}
\left(\tau_{\text{AR}}^{(0)}\right)^{-1} & = & \frac{2\pi}{\hbar Z_{B}^{(0)}}\sum_{bckj}e^{-\beta\left(\varepsilon_{b}-\varepsilon_{j}+\varepsilon_{c}-\varepsilon_{k}\right)}\sum_{a}\left|V_{bacj}\right|^{2}\delta\left(\varepsilon_{b}+\varepsilon_{c}-\varepsilon_{j}-\varepsilon_{a}\right)\nonumber \\
 & + & \frac{2\pi}{\hbar Z_{B}^{(0)}}\sum_{bckj}e^{-\beta\left(\varepsilon_{b}-\varepsilon_{j}+\varepsilon_{c}-\varepsilon_{k}\right)}\sum_{i}\left|V_{ijbk}\right|^{2}\delta\left(\varepsilon_{b}-\varepsilon_{j}-\varepsilon_{k}+\varepsilon_{i}\right).\label{eq:AR NI final}
\end{eqnarray}
The first term on the right hand side (rhs) of equation~\eqref{eq:AR NI final}
describes the decay of a negative trion of energy $\varepsilon_{b}+\varepsilon_{c}-\varepsilon_{j}$
into an electron of energy $\varepsilon_{a}$ while one of the holes
remains a spectator (we refer to this as the ``electron channel''
and it is shown pictorially on the left side of Fig.~\ref{fig:ar-picture}),
and the second term on the rhs of equation~\eqref{eq:AR NI final}
describes the decay of a positive trion of energy $\varepsilon_{b}-\varepsilon_{j}-\varepsilon_{k}$
into a hole of energy $\varepsilon_{i}$ while one of the electrons
remains a spectator (we refer to this as the ``hole channel'').
The explicit form of the Coulomb coupling is then given by:
\begin{eqnarray}
V_{rsut} & = & \iint\frac{\phi_{r}\left(\mathbf{r}\right)\phi_{s}\left(\mathbf{r}\right)\phi_{u}\left(\mathbf{r^{\prime}}\right)\phi_{t}\left(\mathbf{r}^{\prime}\right)}{\left|\mathbf{r}-\mathbf{r}^{\prime}\right|}d^{3}\mathbf{r}\,d^{3}\mathbf{r}^{\prime},
\end{eqnarray}
where $\phi_{s}\left(\mathbf{r}\right)$ are the quasiparticle states
for electrons ($s\in a$) or holes ($s\in i$) and there is no screening
-- consistent with Ref.~\citenum{Wang2003} and Ref.~\citenum{Refaely-Abramson2017}.

As discussed in the introduction, the noninteracting approach is suitable
for nanostructures in the very strong confinement regime, where the
kinetic energy is large compared to electron--hole interactions.
This approach fails, as shown below, for system sizes in the moderate
to weak confinement regimes. The inclusion of electron--hole correlations
is mainly of significance in the description of the initial biexcitonic
states while for the final excitonic states, the noninteracting framework
seems suitable even for weakly confined structures, since the final
state describes a highly excited electron--hole pair, above their
ionization energy. Therefore, we use a noninteracting description
for $\left|S\right\rangle $ given by equation~\eqref{eq:S NI},
but include electron--hole correlations in the description of the
initial biexcitonic state. Motivated by the work of Refaely--Abramson
\emph{et al.,}\cite{Refaely-Abramson2017}\emph{ }we express the biexcitonic
state as two spatially noninteracting but spin--correlated excitons.
This is justified since electron--hole correlations are most significant
within excitons as reflected by the larger exciton binding energy
compared to that of biexcitons.\cite{Patton2003} In our interacting
approach the biexcitonic states take the form:
\begin{eqnarray}
\left|B\right\rangle  & = & \sum_{b,j}\sum_{c,k}c_{b,j}^{B}c_{c,k}^{B}a_{b}^{\dagger}a_{j}a_{c}^{\dagger}a_{k}\left|0\right\rangle \otimes\left|\chi_{B}\right\rangle ,\label{eq:B I}
\end{eqnarray}
where the coefficients $c_{b,j}^{B}$ are determined by solving the
Bethe--Salpeter equation (BSE),\cite{Rohlfing2000} as detailed in
Ref.~\citenum{Eshet2013}. The excitonic energy is given by the noninteracting
expression, while the biexcitonic energy is now a sum of the exciton
energies, each obtained from the BSE. Within the interacting framework,
the AR lifetime is given as a sum of electron--dominated (shown pictorially
on the right side of Fig.~\ref{fig:ar-picture}) and hole--dominated
contributions:
\begin{eqnarray}
\tau_{\text{AR}}^{-1} & =\frac{2\pi}{\hbar Z_{B}} & \sum_{B}e^{-\beta E_{B}}\sum_{a,i}\left|\sum_{b,c,j}c_{b,i}^{B}c_{c,j}^{B}V_{bacj}\right|^{2}\delta\left(E_{B}-\varepsilon_{a}+\varepsilon_{i}\right)\nonumber \\
 & +\frac{2\pi}{\hbar Z_{B}} & \sum_{B}e^{-\beta E_{B}}\sum_{a,i}\left|\sum_{j,b,k}c_{a,j}^{B}c_{b,k}^{B}V_{ijbk}\right|^{2}\delta\left(E_{B}-\varepsilon_{a}+\varepsilon_{i}\right),\label{eq:AR I final}
\end{eqnarray}
where there are \emph{coherent} sums of the Coulomb matrix elements
multiplied with the coefficients that were obtained by diagonalizing
the Bethe--Salpeter Hamiltonian matrix. Due to the presence of electron--hole
interactions, all particles are involved in the AR process in the
interacting formalism. For further details regarding the theory and
the derivations of the above equations, please consult the Methods
section.

\begin{figure}[t]
\centering{}\includegraphics[width=10cm]{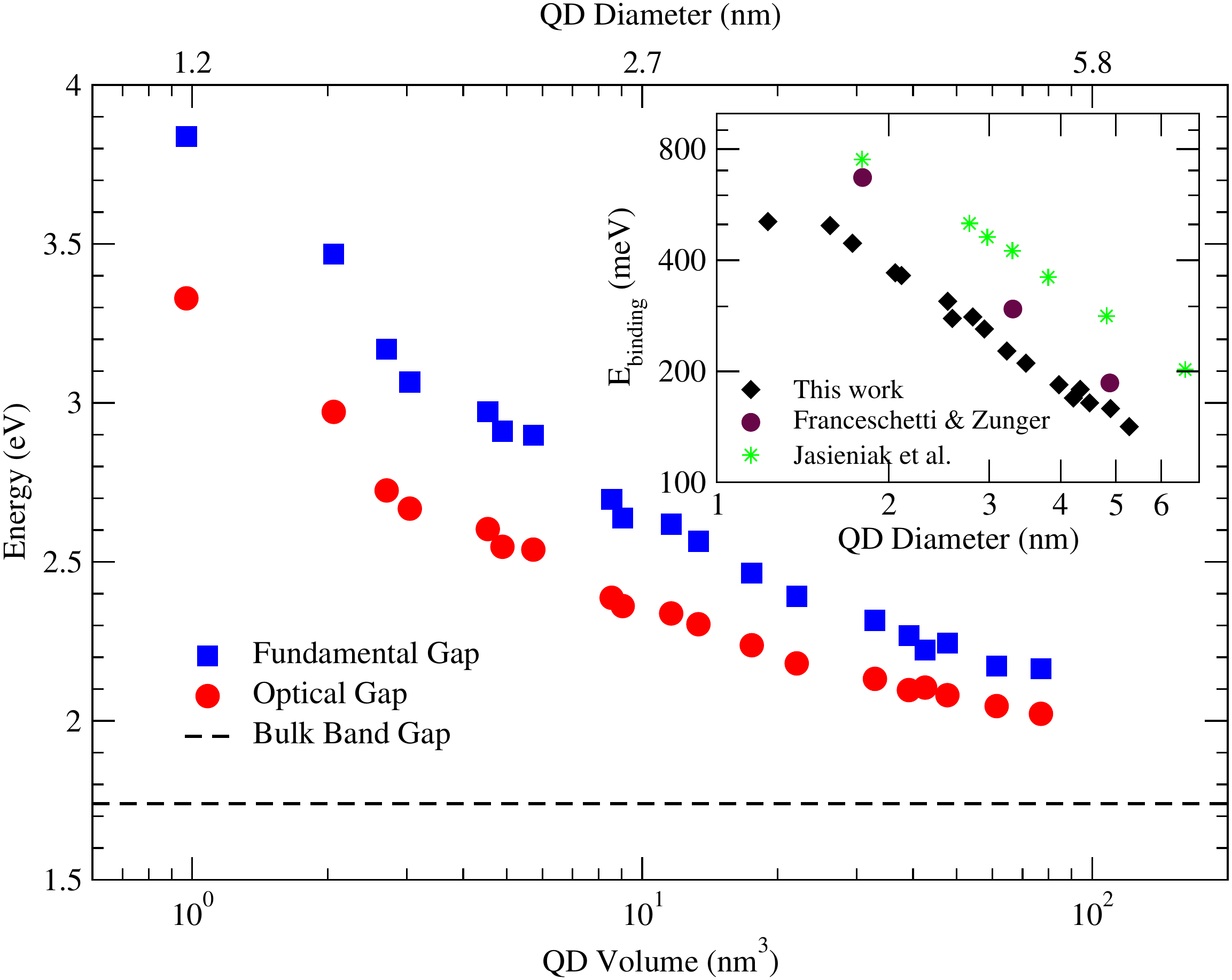}\caption{\label{fig:qd_e_gaps}Energy gaps (in eV) for the seventeen CdSe QDs.
The fundamental gap is shown in blue solid squares and the optical
gap is shown in red solid circles. The inset shows the exciton binding
energy (the energy difference between the fundamental and optical
gaps) which ranges from $\sim500$~meV for the smallest QDs to $\sim150$~meV
for the largest QDs studied here. For comparison, we also show the
measured exciton binding energy (green stars, Ref.~\citenum{Jasieniak2011})
and calculations based on a semi--empirical pseudopotential model
using a perturbative scheme (maroon circles, Ref.~\citenum{Franceschetti1997}).}
\end{figure}

For the implementation of the above frameworks, we chose the semi--empirical
pseudopotential method to model the quasiparticle states.\cite{Wang1994,Wang1996,Rabani1999b,Williamson2000}
And because we only need quasiparticle states in specific energy ranges
(near the band--edge for the initial biexcitonic states and those
that satisfy energy conservation for the final excitonic states),
we utilize the filter--diagonalization technique~\cite{Wall1995,Toledo2002}
to obtain only the required electron and hole eigenstates.\cite{Toledo2002}
Electron--hole correlations were included in the interacting formalism
by solving the BSE within the static screening approximation, where
the dielectric constant was taken from the work of Wang \& Zunger.\cite{Wang1996}

For QDs, we calculated the AR lifetimes for seventeen wurtzite CdSe
QDs with diameters ranging from $D_{\text{QD}}=2R_{\text{QD}}=1.2$~nm
(Cd\textsubscript{20}Se\textsubscript{19}) to $D_{\text{QD}}=2R_{\text{QD}}=5.3$~nm
(Cd\textsubscript{1358}Se\textsubscript{1360}). For completeness,
we also calculated the fundamental and optical gaps for the CdSe QDs,
shown in Fig.~\ref{fig:qd_e_gaps}. The difference in the band and
optical gap is the exciton binding energy and is in good agreement
with previous studies.\cite{Franceschetti1997,Jasieniak2011} This
suggests that (a) our model is accurate enough to reproduce single--
(fundamental gap) and two--particle (optical gap) properties with
the simplification of a uniform dielectric screening and (b) that
our computational machinery shows mild scaling with the system size,
allowing a direct comparison with experiments for realistic NC sizes.

\begin{figure}[t]
\centering{}\includegraphics[width=10cm]{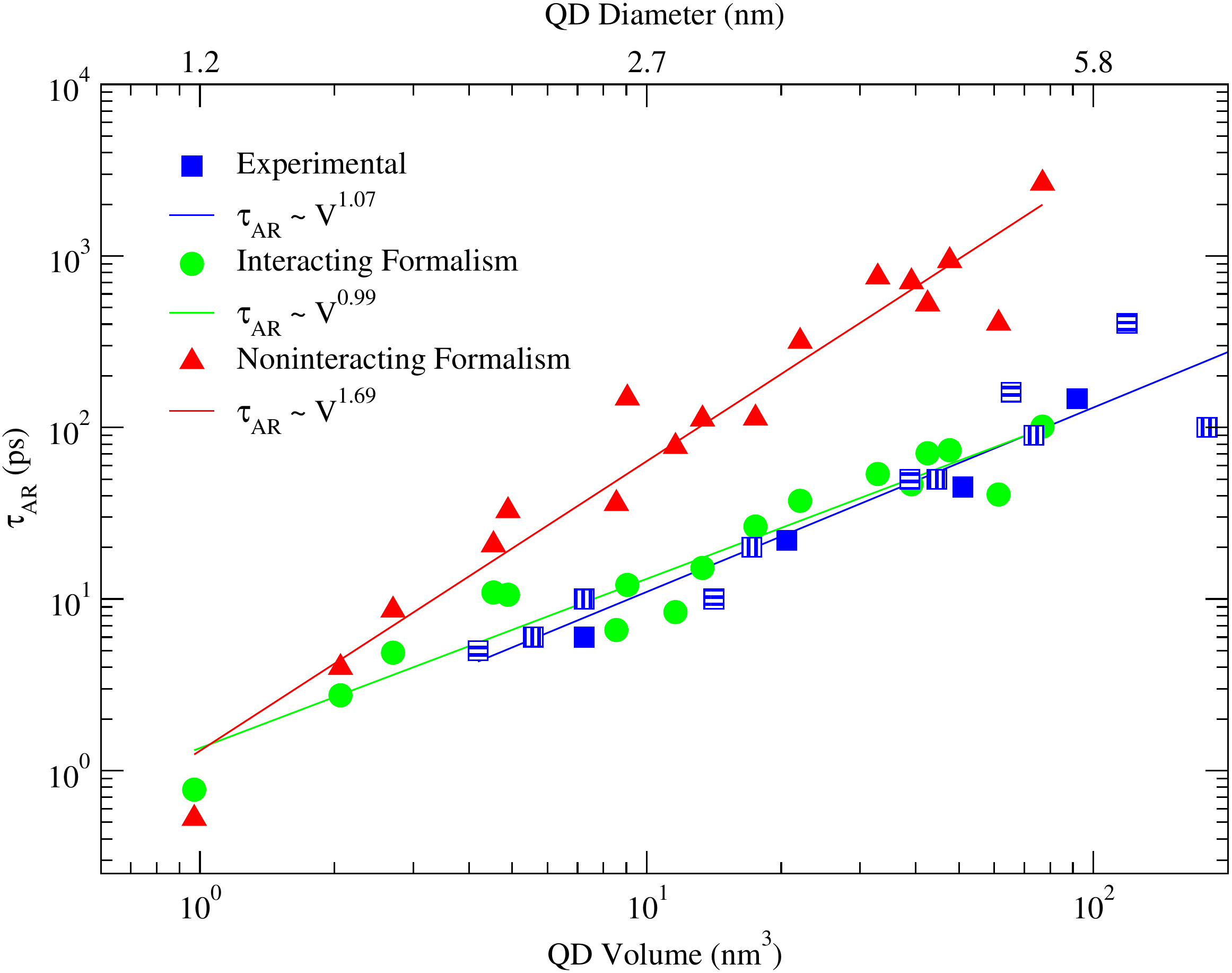}\caption{\label{fig:qd_ar_lifetimes}AR lifetimes, $\tau_{\text{AR}}$, for
CdSe QDs as a function of the volume of the QD. Good agreement is
observed between the interacting formalism (green circles) and experimental
(blue squares: solid,\cite{Klimov2000} vertical lines~\cite{Htoon2003}
and horizontal lines~\cite{Taguchi2011}) AR lifetimes for all sizes.
On the other hand, the noninteracting formalism (red triangles) deviates
from the experimental values for QD volumes $>10$ nm\protect\textsuperscript{3}.
Power law fits, $\tau_{\text{AR}}=a\times V^{b}$, are also shown
for each of the three sets of AR lifetimes.}
\end{figure}

Fig.~\ref{fig:qd_ar_lifetimes} displays the AR lifetimes obtained
by using both the noninteracting (equation~\eqref{eq:AR NI final})
and interacting (equation~\eqref{eq:AR I final}) formalisms along
with experimental~\cite{Klimov2000,Htoon2003,Taguchi2011} measurements
of the AR lifetimes. It is clear that neglecting electron--hole correlations
in the initial biexcitonic state is only reasonable in the very strong
confinement limit, where $R_{\text{QD}}\ll a_{\text{B}}$ (where $a_{\text{B}}=5.6$~nm
is the exciton Bohr radius of CdSe).\cite{Shabaev2004} The noninteracting--based
AR lifetimes increase too rapidly as the volume of the QD increases
compared to both the interacting formalism and experimentally measured
AR lifetimes.
Quantitatively, the computed scaling of the AR lifetime by the noninteracting
formalism is $\tau_{\text{AR,QD}}^{(0)}\propto V^{1.69}$, which is
in contrast to the known volume scaling of the AR lifetime in single
material QDs.\cite{Klimov2000} On the other hand, the volume scaling
is accurately captured by the interacting formalism ($\tau_{\text{AR,QD}}\propto V^{0.99}$),
and the overall agreement with the experiments is remarkable. Recall
that the previous theoretical studies using a noninteracting formalism
for the AR lifetime either studied QDs small enough that the noninteracting
formalism was able to relatively accurately predict the volume scaling
of the AR lifetime~\cite{Wang2003} or the theories predicted a stronger
dependence on the volume ($\propto V^{5/3}$ to $V^{2}$).\cite{Chepic1990,Vaxenburg2015} 

\begin{figure}[t]
\centering{}\includegraphics[width=10cm]{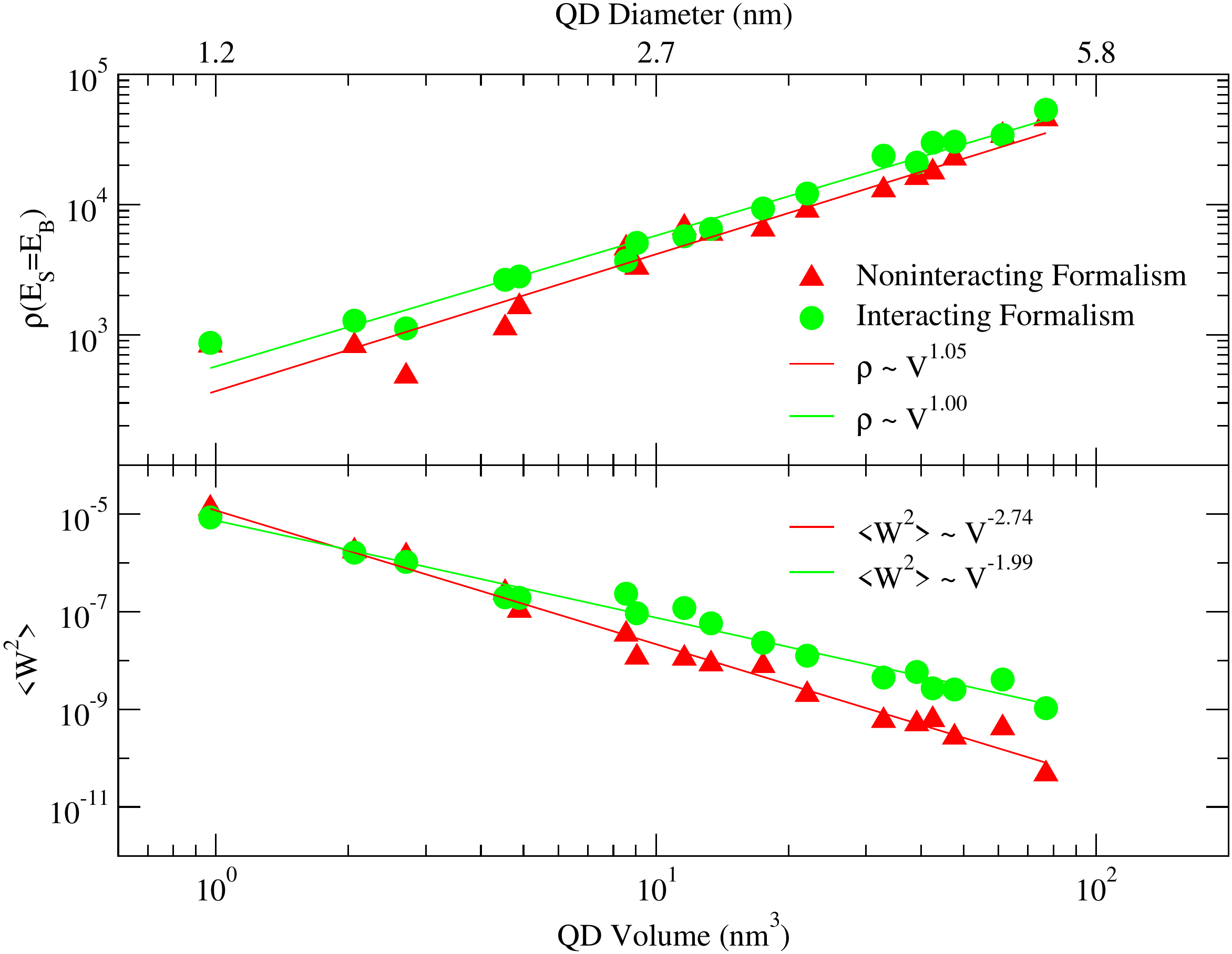}\caption{\label{fig:qd_dos_cc}The top half shows the density of states at
the energy of the hot electron and holes satisfying energy conservation
for CdSe QDs as a function of the volume of the QD. The hot electrons
(holes) have energies approximately $E_{\text{g}}$ above (below)
the HOMO (LUMO) in the noninteracting case and in the interacting
formalism the hot electrons (holes) have energies approximately $E_{\text{opt}}$
above (below) the HOMO (LUMO). The bottom half shows the average of
the Coulomb couplings, $\left\langle W^{2}\right\rangle $, squared
to the final states. The noninteracting formalism results are shown
as red triangles and the interacting formalism results are shown as
green circles. Power law fits, $f\left(V\right)=a\times V^{b}$, are
also shown for all sets.}
\end{figure}

To understand the origin of the volume scaling of the AR lifetimes
for QDs, we start with Fermi's golden rule and, for simplicity, focus
on the rate of decay to hot electrons via the electron channel (similar
arguments also hold for the hole channel) at zero temperature ($b=c\equiv\ell=\text{LUMO}$
and $j\equiv h=\text{HOMO}$) in the noninteracting approach: 
\begin{eqnarray}
\left(\tau_{\text{AR,e}}^{(0)}\right)^{-1} & = & \frac{2\pi}{\hbar}\sum_{a}\left|V_{\ell a\ell h}\right|^{2}\delta\left(\varepsilon_{\ell}+\varepsilon_{\ell}-\varepsilon_{h}-\varepsilon_{a}\right),\label{eq:FGR Averaged}
\end{eqnarray}
where $\varepsilon_{\ell}+\varepsilon_{\ell}-\varepsilon_{h}=2E_{\text{g}}$
equals two times the fundamental gap, $E_{\text{g}}$. The scaling
of the AR lifetime depends on the scaling of the final density of
state and the Coulomb coupling. The former scales linearly with the
volume of the NC.\cite{Rabani2010,Baer2012} Determining the scaling
of the latter is more involved. Naively, one would predict it to scale
with $R_{\text{QD}}^{-1}$ due to the Coulomb potential. However,
because the final hot electron state is highly oscillatory, reflecting
the high kinetic energy of the hot electron, and the initial biexcitonic
state is slowly varying, the leading term that scales as $R_{\text{QD}}^{-1}$
vanishes. The next term, which can be obtained by invoking the stationary
phase approximation, scales as $R_{\text{QD}}^{-3}$.\cite{Chepic1990}
Altogether, these arguments predict an Auger lifetime that is proportional
to the volume: $\tau_{\text{AR,e}}^{-1}\propto\left|R_{\text{QD}}^{-3}\right|^{2}R_{\text{QD}}^{3}\propto R_{\text{QD}}^{-3}$.
Similar arguments hold for the scaling of the Auger lifetime in the
interacting formalism.

We find, as predicted, that the density of hot electrons and holes
scales linearly with the volume of the NCs (top panel, Fig.~\ref{fig:qd_dos_cc})
in both formalisms. However, the scaling of the average Coulomb coupling
squared shows significant deviations from the expected $V^{-2}$ stationary
phase result in the noninteracting formalism ($\propto V^{-2.74}$),
while in the interacting formalism it scales as expected, $\propto V^{-1.99}$.
These different scalings can be rationalized by a more localized electron--hole
wavefunction in the interacting case, due to the screened Coulomb
electron--hole attraction term in the BSE, leading to more overlap
with the wavefunction of the hot electron.

Surprisingly, the noninteracting formalism shows pronounced deviations
from the interacting formalism for CdSe QDs with diameters as small
as $\sim2.5$~nm, much smaller than the exciton Bohr radius ($a_{\text{B}}=5.6$~nm
for CdSe).\cite{Shabaev2004} This was a rather surprising result
as all QDs studied here have $R_{\text{QD}}<a_{\text{B}}$, where
electron--hole interactions are rather small compared to the confinement
kinetic energy (see inset in Fig.~\ref{fig:qd_e_gaps}). 

\begin{figure}[t]
\centering{}\includegraphics[width=10cm]{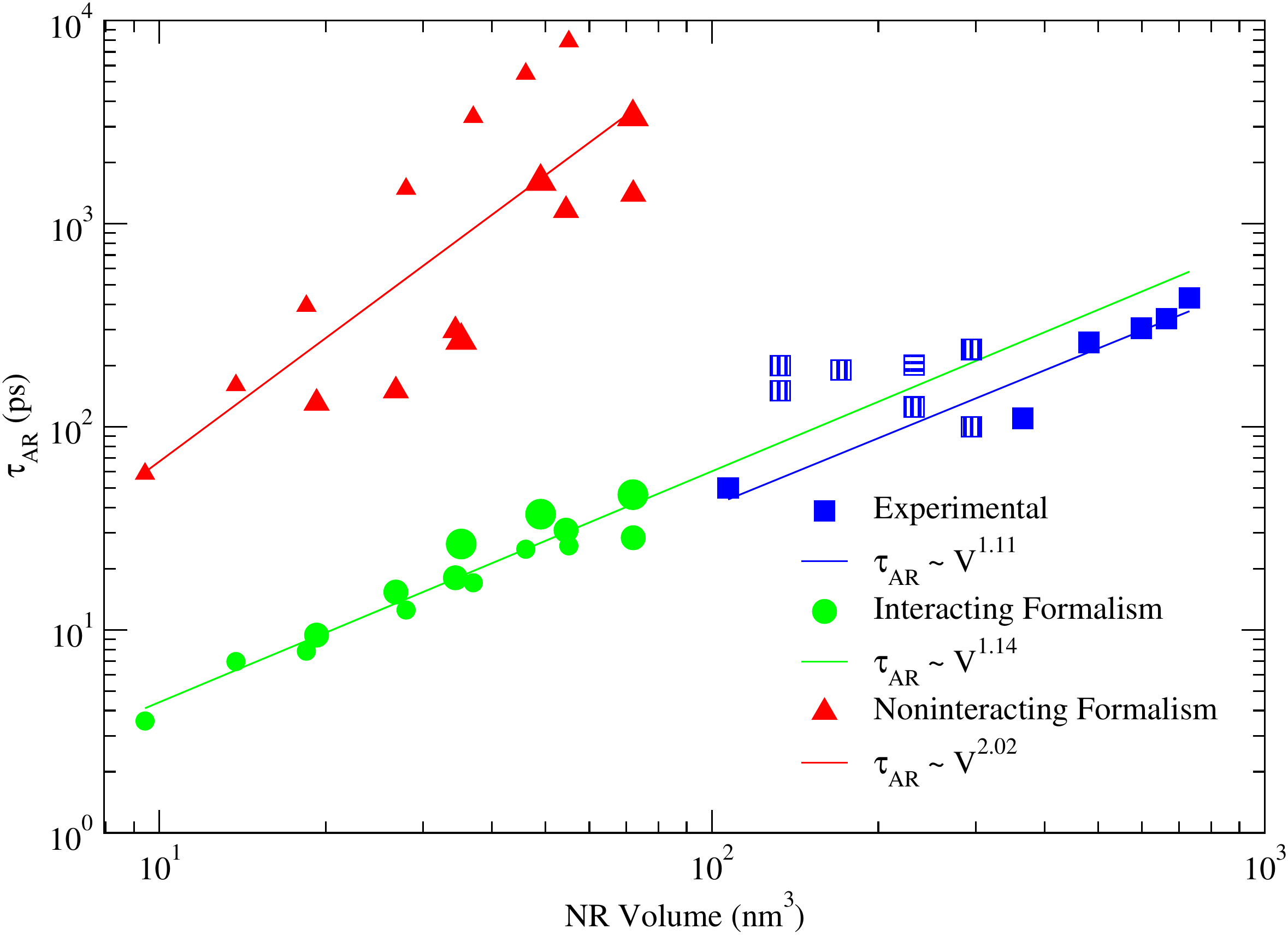}\caption{\label{fig:nr_ar_volume}Auger recombination lifetimes for CdSe NRs
as a function of the volume of the NRs predicted by the interacting
(green circles), the noninteracting (red triangles) formalisms along
with experimentally measured (blue squares: solid~\cite{Htoon2003},
vertical~\cite{Taguchi2011} and horizontal~\cite{Zhu2012} lines)
AR lifetimes. The three different sizes used correspond to the three
different diameters (1.53~nm, 2.14~nm and 2.89~nm) studied computationally.
Power law fits, $\tau_{\text{AR}}=a\times V^{b}$, are also shown
for each of the three sets of AR lifetimes.}
\end{figure}

The deviations in AR lifetimes predicted by the two formalisms are
even larger for CdSe NRs. In Fig.~\ref{fig:nr_ar_volume} we show
the calculated and measured~\cite{Htoon2003,Taguchi2011,Zhu2012}
AR lifetimes for a series of CdSe NRs of different volumes. It is
immediately evident that the noninteracting formalism is quantitatively
incorrect for all NRs studied. The noninteracting--based AR lifetimes
are also too long by approximately $1-2$ orders of magnitude! This
result arises from an underestimation of the Coulomb coupling due
to the electron--hole wavefunctions being delocalized over the entire
NR in the noninteracting formalism; there is no electron--hole attraction
to localize the electron--hole pair to form a bound Wannier exciton
in the noninteracting formalism. In contrast, the interacting formalism
predicts the scaling (nearly linearly with volume) as well as the
magnitude of the AR lifetimes quiet accurately in comparison with
the experimental results depicted by the solid blue squares.\cite{Htoon2003}
Based on the results reported for spherical QDs, this is to be expected
and further signifies the importance of electron--hole correlations
in the AR process in confined nanostructures. 

Interestingly, more recent experimental measurements show nearly no
volume effect on the AR lifetimes in CdSe NRs (striped blue square),\cite{Taguchi2011}
however, the same authors reported on the inconsistencies between
transient absorption and time--resolved photoluminescence measurements
(for the largest system studied, the two measurements differ by a
factor of $\approx3$). Similar inconsistencies for NRs were reported
for the reverse process, by which a hot exciton decays into a biexcitonic
state by impact excitation, leading to multiexciton generation (MEG).
Preliminary measurements reported a notable volume dependence of the
impact excitation rate,\cite{Cunningham2011,Sandberg2012} while more
recent theoretical work,\cite{Baer2013} followed by experimental
validation,\cite{Padilha2013} argued that impact excitation rates
are volume independent. This suggests that different experimental
setups (synthesis and optical measurements) may lead to different
scaling behavior. A similar reasoning may also explain the discrepancy
between the two sets of experimental results on AR lifetimes shown
in Fig.~\ref{fig:nr_ar_volume}. However, more experimental work
is needed to fully understand the diversity of experimental outcomes,
in particular, given that our new theoretical predictions are consistent
with one set of measurements but not the other.

Returning to the AR lifetime scaling with volume in NRs, the noninteracting
formalism behaves as $\tau_{\text{AR,NR}}^{(0)}\propto V^{2.02}$.
This is expected based on the scaling of the Coulomb matrix elements
with the diameter and length of the NR,\cite{Baer2013} but is in
contrast to the scaling observed both experimentally~\cite{Htoon2003}
and theoretically using the interacting formalism. Thus, including
electron--hole correlations is needed for both a quantitatively and
qualitatively accurate description of the AR lifetime calculation
in NRs. Intuitively, this result makes sense due to both the lack
of confinement along the NR axis and the large electron--hole binding
energy in CdSe NRs ($\sim200$~meV)~\cite{Shabaev2004} contributing
to making the noninteracting carrier approximation invalid in NRs.

\begin{figure}[t]
\begin{centering}
\includegraphics[width=12cm]{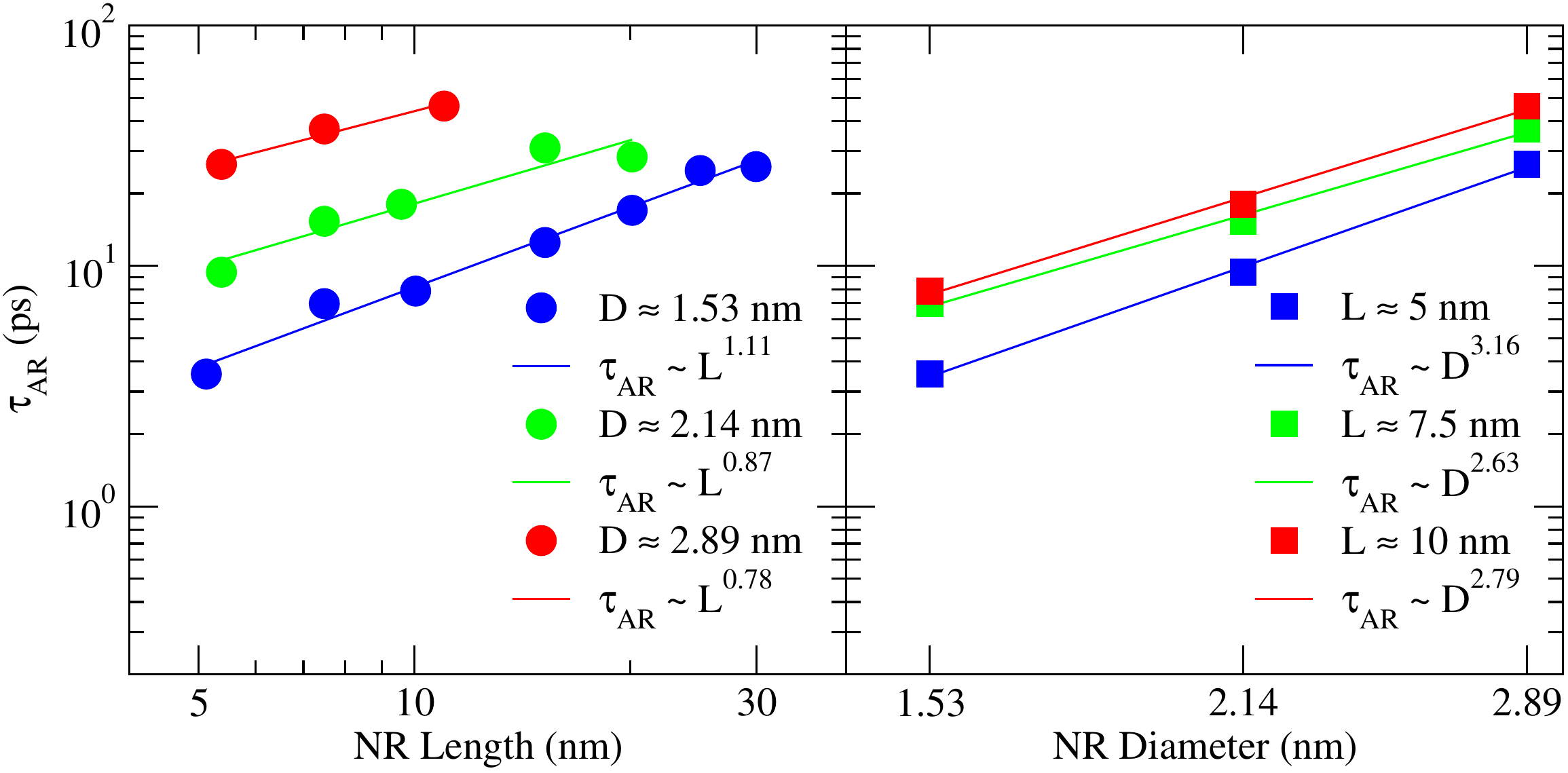}
\par\end{centering}
\caption{\label{nr_ar_len_diam}Interacting formalism based Auger recombination
lifetimes for CdSe NRs as a function of the length (left) and diameter
(right) of the NR. Power law fits, $\tau_{\text{AR}}=a\times D^{b}$
and $\tau_{\text{AR}}=a\times L^{b}$, are also shown for each NR
set.}
\end{figure}

As mentioned above, it is experimentally difficult to independently
control the NR diameter and length; however, it is trivial to do computationally,
so we analyzed the AR lifetime scaling separately for the NR diameter
and length. We found that the AR lifetime scales approximately \emph{quadratically--cubically}\textbf{
}with the length of the NR in the noninteracting formalism, while
it scales nearly \emph{linearly} in the interacting formalism (Fig.~\ref{nr_ar_len_diam}),
in agreement with previous experimental measurements.\cite{Htoon2003,Yang2012,Padilha2013,Aerts2013,Stolle2017}
However, the scaling with the length of the NR depends slightly on
the diameter. We also observed an approximate $D^{3}$\emph{ }scaling
in the interacting formalism, which still awaits experimental validation.

Our finding that the noninteracting formalism is inaccurate for NRs
whereas the interacting formalism is accurate further corroborates
previous kinetic models and experiments that argued that the total
AR rate in NRs increases quadratically with the number of excitons,
$n$ $\left(k_{\text{AR}}\left(n\right)\propto n\left(n-1\right)/2\right)$.\cite{Barzykin2005,Zhu2012,Aerts2013,Ben-Shahar2018}
In other words, kinetic models of AR in NRs should model AR as a bimolecular
collision of two excitons; in opposition to the combinatorial scaling
of $n^{2}\left(n-1\right)/2$ if modeling AR as a three particle collision
between free, noninteracting electrons and holes. Overall, these results
on CdSe NRs add to the body of work that electrons and holes form
bound 1D Wannier excitons in 1D systems such as semiconductor NRs
and carbon nanotubes.\cite{Ma2005,Huang2006,Wang2011,Pal2017}

In conclusion, the interacting approach developed here for calculating
AR lifetimes in NCs provides a framework that is able to predict quantitatively
accurate AR lifetimes in both QDs and NRs. Our interacting formalism
is the first to postdict the experimentally observed linear volume
dependence of the AR lifetime in QDs as well as the correct scaling
of the AR lifetimes in NRs with respect to the length and volume.
This result was rationalized by noting that the matrix elements in
AR lifetime calculations involve a product of the initial electron
and hole states; thus, taking into account electron--hole correlations
will have a large impact in regimes where the confinement energy is
comparable or smaller than the exciton binding energy. Electron--hole
correlations result in a localization of the pair, thereby, increasing
the Coulomb coupling between the initial and final states. This is
especially true in NRs where the lack of confinement along the NR
axis makes the electron--hole attraction even more important. The
resulting localization of the electron--hole pair leads to dramatic
decreases in the AR lifetimes, as large as $2$ orders of magnitude,
when including such correlations. 

Altogether, the interacting formalism outlined in this Letter constitutes
a large step in bringing theoretical studies up to speed with ability
of experimentalists to measure AR lifetimes and, in general, multiexciton
dynamics. Our approach allows for direct comparisons and joint investigations
between theorists and experimentalists as it permits accurate theoretical
calculations of AR lifetimes for experimentally relevant nanostructures
of any dimensionality and composition. It should be noted that our
framework assumes the excitons scatter coherently; thus, systems in
which exciton diffusion is the rate limiting step are currently outside
the scope of our approach. In future work we plan to apply our formalism
to study AR in CdSe NPLs and extend it to also include exciton diffusion
processes, to resolve another experimentally controversy where two
different methods provide significantly different scaling behaviors
in 2D NPLs.\cite{She2015,Li2017} 
\begin{methods}
A detailed derivation of the equations along with additional information
and discussion on the implementation of the theory using the semi--empirical
pseudopotential method, filter--diagonalization technique, Bethe--Salpeter
equation, Fermi's golden rule in the AR lifetime calculations presented
in this Letter and the procedure used to construct the CdSe QDs and
NRs is also outlined. This material is available at https://doi.org/.
\end{methods}

\bibliography{auger_qds_nrs_manuscript}

\begin{thebibliography}{10}
\expandafter\ifx\csname url\endcsname\relax
  \def\url#1{\texttt{#1}}\fi
\expandafter\ifx\csname urlprefix\endcsname\relax\def\urlprefix{URL }\fi
\providecommand{\bibinfo}[2]{#2}
\providecommand{\eprint}[2][]{\url{#2}}

\bibitem{Klimov2000}
\bibinfo{author}{Klimov, V.~I.}, \bibinfo{author}{Mikhailovsky, A.~A.},
  \bibinfo{author}{McBranch, D.~W.}, \bibinfo{author}{Leatherdale, C.~A.} \&
  \bibinfo{author}{Bawendi, M.~G.}
\newblock \bibinfo{title}{{Quantization of Multiparticle Auger Rates in
  Semiconductor Quantum Dots}}.
\newblock \emph{\bibinfo{journal}{Science}} \textbf{\bibinfo{volume}{287}},
  \bibinfo{pages}{1011--1013} (\bibinfo{year}{2000}).

\bibitem{Klimov2014}
\bibinfo{author}{Klimov, V.~I.}
\newblock \bibinfo{title}{{Multicarrier Interactions in Semiconductor
  Nanocrystals in Relation to the Phenomena of Auger Recombination and Carrier
  Multiplication}}.
\newblock \emph{\bibinfo{journal}{Annu. Rev. Condens. Matter Phys.}}
  \textbf{\bibinfo{volume}{5}}, \bibinfo{pages}{285--316}
  (\bibinfo{year}{2014}).

\bibitem{Colvin1994}
\bibinfo{author}{Colvin, V.~L.}, \bibinfo{author}{Schlamp, M.~C.} \&
  \bibinfo{author}{Alivisatos, A.~P.}
\newblock \bibinfo{title}{{Light-emitting diodes made from cadmium selenide
  nanocrystals and a semiconducting polymer}}.
\newblock \emph{\bibinfo{journal}{Nature}} \textbf{\bibinfo{volume}{370}},
  \bibinfo{pages}{354--357} (\bibinfo{year}{1994}).

\bibitem{Imamoglu2000}
\bibinfo{author}{Imamoglu, A.} \emph{et~al.}
\newblock \bibinfo{title}{{Quantum correlation among photons from a single
  quantum dot at room temperature}}.
\newblock \emph{\bibinfo{journal}{Nature}} \textbf{\bibinfo{volume}{406}},
  \bibinfo{pages}{968--970} (\bibinfo{year}{2000}).

\bibitem{Pietryga2008}
\bibinfo{author}{Pietryga, J.~M.}, \bibinfo{author}{Zhuravlev, K.~K.},
  \bibinfo{author}{Whitehead, M.}, \bibinfo{author}{Klimov, V.~I.} \&
  \bibinfo{author}{Schaller, R.~D.}
\newblock \bibinfo{title}{{Evidence for barrierless auger recombination in PbSe
  nanocrystals: A pressure-dependent study of transient optical absorption}}.
\newblock \emph{\bibinfo{journal}{Phys. Rev. Lett.}}
  \textbf{\bibinfo{volume}{101}}, \bibinfo{pages}{217401}
  (\bibinfo{year}{2008}).

\bibitem{Bae2013}
\bibinfo{author}{Bae, W.~K.} \emph{et~al.}
\newblock \bibinfo{title}{{Controlled alloying of the core-shell interface in
  CdSe/CdS quantum dots for suppression of auger recombination}}.
\newblock \emph{\bibinfo{journal}{ACS Nano}} \textbf{\bibinfo{volume}{7}},
  \bibinfo{pages}{3411--3419} (\bibinfo{year}{2013}).

\bibitem{Fathpour2004}
\bibinfo{author}{Fathpour, S.} \emph{et~al.}
\newblock \bibinfo{title}{{The role of Auger recombination in the
  temperature-dependent output characteristics (T$_0$ = $\infty$) of p-doped
  1.3 $\mu$m quantum dot lasers}}.
\newblock \emph{\bibinfo{journal}{Appl. Phys. Lett.}}
  \textbf{\bibinfo{volume}{85}}, \bibinfo{pages}{5164--5166}
  (\bibinfo{year}{2004}).

\bibitem{Sukhovatkin2009}
\bibinfo{author}{Sukhovatkin, V.}, \bibinfo{author}{Hinds, S.},
  \bibinfo{author}{Brzozowski, L.} \& \bibinfo{author}{Sargent, E.~H.}
\newblock \bibinfo{title}{{Colloidal quantum-dot photodetectors exploiting
  multiexciton generation}}.
\newblock \emph{\bibinfo{journal}{Science}} \textbf{\bibinfo{volume}{324}},
  \bibinfo{pages}{1542--1544} (\bibinfo{year}{2009}).

\bibitem{Nair2011}
\bibinfo{author}{Nair, G.}, \bibinfo{author}{Zhao, J.} \&
  \bibinfo{author}{Bawendi, M.~G.}
\newblock \bibinfo{title}{{Biexciton quantum yield of single semiconductor
  nanocrystals from photon statistics}}.
\newblock \emph{\bibinfo{journal}{Nano Lett.}} \textbf{\bibinfo{volume}{11}},
  \bibinfo{pages}{1136--1140} (\bibinfo{year}{2011}).

\bibitem{Ben-Shahar2018}
\bibinfo{author}{Ben-Shahar, Y.} \emph{et~al.}
\newblock \bibinfo{title}{{Charge carrier dynamics in photocatalytic hybrid
  semiconductor-metal nanorods: crossover from Auger recombination to charge
  transfer.}}
\newblock \emph{\bibinfo{journal}{Nano Lett.}} \textbf{\bibinfo{volume}{18}},
  \bibinfo{pages}{5211--5216} (\bibinfo{year}{2018}).

\bibitem{Robel2009}
\bibinfo{author}{Robel, I.}, \bibinfo{author}{Gresback, R.},
  \bibinfo{author}{Kortshagen, U.}, \bibinfo{author}{Schaller, R.~D.} \&
  \bibinfo{author}{Klimov, V.~I.}
\newblock \bibinfo{title}{{Universal size-dependent trend in auger
  recombination in direct-gap and indirect-gap semiconductor nanocrystals}}.
\newblock \emph{\bibinfo{journal}{Phys. Rev. Lett.}}
  \textbf{\bibinfo{volume}{102}}, \bibinfo{pages}{177404}
  (\bibinfo{year}{2009}).

\bibitem{Garcia-Santamaria2009}
\bibinfo{author}{Garc{\'{i}}a-Santamar{\'{i}}a, F.} \emph{et~al.}
\newblock \bibinfo{title}{{Suppressed auger recombination in "Giant"
  nanocrystals boosts optical gain performance}}.
\newblock \emph{\bibinfo{journal}{Nano Lett.}} \textbf{\bibinfo{volume}{9}},
  \bibinfo{pages}{3482--3488} (\bibinfo{year}{2009}).

\bibitem{Pijpers2010}
\bibinfo{author}{Pijpers, J. J.~H.}, \bibinfo{author}{Milder, M. T.~W.},
  \bibinfo{author}{Delerue, C.} \& \bibinfo{author}{Bonn, M.}
\newblock \bibinfo{title}{{(Multi)exciton Dynamics and Exciton Polarizability
  in Colloidal InAs Quantum Dots}}.
\newblock \emph{\bibinfo{journal}{J. Phys. Chem. C.}}
  \textbf{\bibinfo{volume}{114}}, \bibinfo{pages}{6318--6324}
  (\bibinfo{year}{2010}).

\bibitem{Chepic1990}
\bibinfo{author}{Chepic, D.~I.} \emph{et~al.}
\newblock \bibinfo{title}{{Auger ionization of semiconductor quantum drops in a
  glass matrix}}.
\newblock \emph{\bibinfo{journal}{J. Lumin.}} \textbf{\bibinfo{volume}{47}},
  \bibinfo{pages}{113--127} (\bibinfo{year}{1990}).

\bibitem{Vaxenburg2015}
\bibinfo{author}{Vaxenburg, R.}, \bibinfo{author}{Rodina, A.},
  \bibinfo{author}{Shabaev, A.}, \bibinfo{author}{Lifshitz, E.} \&
  \bibinfo{author}{Efros, A.~L.}
\newblock \bibinfo{title}{{Nonradiative auger recombination in semiconductor
  nanocrystals}}.
\newblock \emph{\bibinfo{journal}{Nano Lett.}} \textbf{\bibinfo{volume}{15}},
  \bibinfo{pages}{2092--2098} (\bibinfo{year}{2015}).

\bibitem{Vaxenburg2016}
\bibinfo{author}{Vaxenburg, R.}, \bibinfo{author}{Rodina, A.},
  \bibinfo{author}{Lifshitz, E.} \& \bibinfo{author}{Efros, A.~L.}
\newblock \bibinfo{title}{{Biexciton Auger Recombination in CdSe/CdS Core/Shell
  Semiconductor Nanocrystals}}.
\newblock \emph{\bibinfo{journal}{Nano Lett.}} \textbf{\bibinfo{volume}{16}},
  \bibinfo{pages}{2503--2511} (\bibinfo{year}{2016}).

\bibitem{Cragg2010}
\bibinfo{author}{Cragg, G.~E.} \& \bibinfo{author}{Efros, A.~L.}
\newblock \bibinfo{title}{{Suppression of auger processes in confined
  structures}}.
\newblock \emph{\bibinfo{journal}{Nano Lett.}} \textbf{\bibinfo{volume}{10}},
  \bibinfo{pages}{313--317} (\bibinfo{year}{2010}).

\bibitem{Zavelani-Rossi2010}
\bibinfo{author}{Zavelani-Rossi, M.}, \bibinfo{author}{Lupo, M.~G.},
  \bibinfo{author}{Tassone, F.}, \bibinfo{author}{Manna, L.} \&
  \bibinfo{author}{Lanzani, G.}
\newblock \bibinfo{title}{{Suppression of biexciton auger recombination in
  CdSe/CdS Dot/rods: Role of the electronic structure in the carrier
  dynamics}}.
\newblock \emph{\bibinfo{journal}{Nano Lett.}} \textbf{\bibinfo{volume}{10}},
  \bibinfo{pages}{3142--3150} (\bibinfo{year}{2010}).

\bibitem{Garcia-Santamaria2011}
\bibinfo{author}{Garc{\'{i}}a-Santamar{\'{i}}a, F.} \emph{et~al.}
\newblock \bibinfo{title}{{Breakdown of volume scaling in auger recombination
  in CdSe/CdS heteronanocrystals: The role of the core-shell interface}}.
\newblock \emph{\bibinfo{journal}{Nano Lett.}} \textbf{\bibinfo{volume}{11}},
  \bibinfo{pages}{687--693} (\bibinfo{year}{2011}).

\bibitem{Park2014}
\bibinfo{author}{Park, Y.~S.}, \bibinfo{author}{Bae, W.~K.},
  \bibinfo{author}{Padilha, L.~A.}, \bibinfo{author}{Pietryga, J.~M.} \&
  \bibinfo{author}{Klimov, V.~I.}
\newblock \bibinfo{title}{{Effect of the core/shell interface on auger
  recombination evaluated by single-quantum-dot spectroscopy}}.
\newblock \emph{\bibinfo{journal}{Nano Lett.}} \textbf{\bibinfo{volume}{14}},
  \bibinfo{pages}{396--402} (\bibinfo{year}{2014}).

\bibitem{Jain2016}
\bibinfo{author}{Jain, A.} \emph{et~al.}
\newblock \bibinfo{title}{{Atomistic Design of CdSe/CdS Core-Shell Quantum Dots
  with Suppressed Auger Recombination}}.
\newblock \emph{\bibinfo{journal}{Nano Lett.}} \textbf{\bibinfo{volume}{16}},
  \bibinfo{pages}{6491--6496} (\bibinfo{year}{2016}).

\bibitem{Jain2017}
\bibinfo{author}{Jain, A.}, \bibinfo{author}{Voznyy, O.},
  \bibinfo{author}{Korkusinski, M.}, \bibinfo{author}{Hawrylak, P.} \&
  \bibinfo{author}{Sargent, E.~H.}
\newblock \bibinfo{title}{{Ultrafast Carrier Trapping in Thick-Shell Colloidal
  Quantum Dots}}.
\newblock \emph{\bibinfo{journal}{J. Phys. Chem. Lett.}}
  \textbf{\bibinfo{volume}{8}}, \bibinfo{pages}{3179--3184}
  (\bibinfo{year}{2017}).

\bibitem{Htoon2003}
\bibinfo{author}{Htoon, H.}, \bibinfo{author}{Hollingsworth, J.~a.},
  \bibinfo{author}{Dickerson, R.} \& \bibinfo{author}{Klimov, V.~I.}
\newblock \bibinfo{title}{{Effect of zero- to one-dimensional transformation on
  multiparticle Auger recombination in semiconductor quantum rods.}}
\newblock \emph{\bibinfo{journal}{Phys. Rev. Lett.}}
  \textbf{\bibinfo{volume}{91}}, \bibinfo{pages}{227401}
  (\bibinfo{year}{2003}).

\bibitem{Taguchi2011}
\bibinfo{author}{Taguchi, S.}, \bibinfo{author}{Saruyama, M.},
  \bibinfo{author}{Teranishi, T.} \& \bibinfo{author}{Kanemitsu, Y.}
\newblock \bibinfo{title}{{Quantized Auger recombination of biexcitons in CdSe
  nanorods studied by time-resolved photoluminescence and transient-absorption
  spectroscopy}}.
\newblock \emph{\bibinfo{journal}{Phys. Rev. B}} \textbf{\bibinfo{volume}{83}},
  \bibinfo{pages}{155324} (\bibinfo{year}{2011}).

\bibitem{Zhu2012}
\bibinfo{author}{Zhu, H.} \& \bibinfo{author}{Lian, T.}
\newblock \bibinfo{title}{{Enhanced multiple exciton dissociation from CdSe
  quantum rods: The effect of nanocrystal shape}}.
\newblock \emph{\bibinfo{journal}{J. Am. Chem. Soc.}}
  \textbf{\bibinfo{volume}{134}}, \bibinfo{pages}{11289--11297}
  (\bibinfo{year}{2012}).

\bibitem{Yang2012}
\bibinfo{author}{Yang, J.}, \bibinfo{author}{Hyun, B.~R.},
  \bibinfo{author}{Basile, A.~J.} \& \bibinfo{author}{Wise, F.~W.}
\newblock \bibinfo{title}{{Exciton relaxation in PbSe nanorods}}.
\newblock \emph{\bibinfo{journal}{ACS Nano}} \textbf{\bibinfo{volume}{6}},
  \bibinfo{pages}{8120--8127} (\bibinfo{year}{2012}).

\bibitem{Padilha2013}
\bibinfo{author}{Padilha, L.~A.} \emph{et~al.}
\newblock \bibinfo{title}{{Aspect Ratio Dependence of Auger Recombination and
  Carrier Multiplication in PbSe Nanorods}}.
\newblock \emph{\bibinfo{journal}{Nano Lett.}} \textbf{\bibinfo{volume}{13}},
  \bibinfo{pages}{1092--1099} (\bibinfo{year}{2013}).

\bibitem{Aerts2013}
\bibinfo{author}{Aerts, M.} \emph{et~al.}
\newblock \bibinfo{title}{{Cooling and auger recombination of charges in PbSe
  nanorods: Crossover from cubic to bimolecular decay}}.
\newblock \emph{\bibinfo{journal}{Nano Lett.}} \textbf{\bibinfo{volume}{13}},
  \bibinfo{pages}{4380--4386} (\bibinfo{year}{2013}).

\bibitem{She2015}
\bibinfo{author}{She, C.} \emph{et~al.}
\newblock \bibinfo{title}{{Red, Yellow, Green, and Blue Amplified Spontaneous
  Emission and Lasing Using Colloidal CdSe Nanoplatelets}}.
\newblock \emph{\bibinfo{journal}{ACS Nano}} \textbf{\bibinfo{volume}{9}},
  \bibinfo{pages}{9475--9485} (\bibinfo{year}{2015}).

\bibitem{Stolle2017}
\bibinfo{author}{Stolle, C.~J.}, \bibinfo{author}{Lu, X.}, \bibinfo{author}{Yu,
  Y.}, \bibinfo{author}{Schaller, R.~D.} \& \bibinfo{author}{Korgel, B.~A.}
\newblock \bibinfo{title}{{Efficient Carrier Multiplication in Colloidal
  Silicon Nanorods}}.
\newblock \emph{\bibinfo{journal}{Nano Lett.}} \textbf{\bibinfo{volume}{17}},
  \bibinfo{pages}{5580--5586} (\bibinfo{year}{2017}).

\bibitem{Li2017}
\bibinfo{author}{Li, Q.} \& \bibinfo{author}{Lian, T.}
\newblock \bibinfo{title}{{Area- and Thickness-Dependent Biexciton Auger
  Recombination in Colloidal CdSe Nanoplatelets: Breaking the Universal Volume
  Scaling Law}}.
\newblock \emph{\bibinfo{journal}{Nano Lett.}} \textbf{\bibinfo{volume}{17}},
  \bibinfo{pages}{3152--3158} (\bibinfo{year}{2017}).

\bibitem{Wang2006}
\bibinfo{author}{Wang, F.}, \bibinfo{author}{Wu, Y.},
  \bibinfo{author}{Hybertsen, M.~S.} \& \bibinfo{author}{Heinz, T.~F.}
\newblock \bibinfo{title}{{Auger recombination of excitons in one-dimensional
  systems}}.
\newblock \emph{\bibinfo{journal}{Phys. Rev. B}} \textbf{\bibinfo{volume}{73}},
  \bibinfo{pages}{245424} (\bibinfo{year}{2006}).

\bibitem{Wang2003}
\bibinfo{author}{Wang, L.-W.}, \bibinfo{author}{Califano, M.},
  \bibinfo{author}{Zunger, A.} \& \bibinfo{author}{Franceschetti, A.}
\newblock \bibinfo{title}{{Pseudopotential theory of Auger processes in CdSe
  quantum dots.}}
\newblock \emph{\bibinfo{journal}{Phys. Rev. Lett.}}
  \textbf{\bibinfo{volume}{91}}, \bibinfo{pages}{056404}
  (\bibinfo{year}{2003}).

\bibitem{Korkusinski2011}
\bibinfo{author}{Korkusinski, M.}, \bibinfo{author}{Voznyy, O.} \&
  \bibinfo{author}{Hawrylak, P.}
\newblock \bibinfo{title}{{Theory of highly excited semiconductor
  nanostructures including Auger coupling: Exciton-biexciton mixing in CdSe
  nanocrystals}}.
\newblock \emph{\bibinfo{journal}{Phys. Rev. B}} \textbf{\bibinfo{volume}{84}},
  \bibinfo{pages}{155327} (\bibinfo{year}{2011}).

\bibitem{Patton2003}
\bibinfo{author}{Patton, B.}, \bibinfo{author}{Langbein, W.} \&
  \bibinfo{author}{Woggon, U.}
\newblock \bibinfo{title}{{Trion, biexciton, and exciton dynamics in single
  self-assembled CdSe quantum dots}}.
\newblock \emph{\bibinfo{journal}{Phys. Rev. B}} \textbf{\bibinfo{volume}{68}},
  \bibinfo{pages}{125316} (\bibinfo{year}{2003}).

\bibitem{Balan2017}
\bibinfo{author}{Balan, A.~D.} \emph{et~al.}
\newblock \bibinfo{title}{{Effect of Thermal Fluctuations on the Radiative Rate
  in Core/Shell Quantum Dots}}.
\newblock \emph{\bibinfo{journal}{Nano Lett.}} \textbf{\bibinfo{volume}{17}},
  \bibinfo{pages}{1629--1636} (\bibinfo{year}{2017}).

\bibitem{Refaely-Abramson2017}
\bibinfo{author}{Refaely-Abramson, S.}, \bibinfo{author}{{Da Jornada}, F.~H.},
  \bibinfo{author}{Louie, S.~G.} \& \bibinfo{author}{Neaton, J.~B.}
\newblock \bibinfo{title}{{Origins of Singlet Fission in Solid Pentacene from
  an ab initio Green's Function Approach}}.
\newblock \emph{\bibinfo{journal}{Phys. Rev. Lett.}}
  \textbf{\bibinfo{volume}{119}}, \bibinfo{pages}{267401}
  (\bibinfo{year}{2017}).

\bibitem{Rohlfing2000}
\bibinfo{author}{Rohlfing, M.} \& \bibinfo{author}{Louie, S.~G.}
\newblock \bibinfo{title}{{Electron-hole excitations and optical spectra from
  first principles}}.
\newblock \emph{\bibinfo{journal}{Phys. Rev. B}} \textbf{\bibinfo{volume}{62}},
  \bibinfo{pages}{4927--4944} (\bibinfo{year}{2000}).

\bibitem{Eshet2013}
\bibinfo{author}{Eshet, H.}, \bibinfo{author}{Gr{\"{u}}nwald, M.} \&
  \bibinfo{author}{Rabani, E.}
\newblock \bibinfo{title}{{The electronic structure of CdSe/CdS Core/shell
  seeded nanorods: Type-I or quasi-type-II?}}
\newblock \emph{\bibinfo{journal}{Nano Lett.}} \textbf{\bibinfo{volume}{13}},
  \bibinfo{pages}{5880--5885} (\bibinfo{year}{2013}).

\bibitem{Jasieniak2011}
\bibinfo{author}{Jasieniak, J.}, \bibinfo{author}{Califano, M.} \&
  \bibinfo{author}{Watkins, S.~E.}
\newblock \bibinfo{title}{{Size-dependent valence and conduction band-edge
  energies of semiconductor nanocrystals}}.
\newblock \emph{\bibinfo{journal}{ACS Nano}} \textbf{\bibinfo{volume}{5}},
  \bibinfo{pages}{5888--5902} (\bibinfo{year}{2011}).

\bibitem{Franceschetti1997}
\bibinfo{author}{Franceschetti, A.} \& \bibinfo{author}{Zunger, A.}
\newblock \bibinfo{title}{{Direct pseudopotential calculation of exciton
  coulomb and exchange energies in semiconductor quantum dots}}.
\newblock \emph{\bibinfo{journal}{Phys. Rev. Lett.}}
  \textbf{\bibinfo{volume}{78}}, \bibinfo{pages}{915--918}
  (\bibinfo{year}{1997}).

\bibitem{Wang1994}
\bibinfo{author}{Wang, L.~W.} \& \bibinfo{author}{Zunger, A.}
\newblock \bibinfo{title}{{Electronic Structure Pseudopotential Calculations of
  Large (.apprx.1000 Atoms) Si Quantum Dots}}.
\newblock \emph{\bibinfo{journal}{J. Phys. Chem.}}
  \textbf{\bibinfo{volume}{98}}, \bibinfo{pages}{2158--2165}
  (\bibinfo{year}{1994}).

\bibitem{Wang1996}
\bibinfo{author}{Wang, L.-W.} \& \bibinfo{author}{Zunger, A.}
\newblock \bibinfo{title}{{Pseudopotential calculations of nanoscale CdSe
  quantum dots}}.
\newblock \emph{\bibinfo{journal}{Phys. Rev. B}} \textbf{\bibinfo{volume}{53}},
  \bibinfo{pages}{9579--9582} (\bibinfo{year}{1996}).

\bibitem{Rabani1999b}
\bibinfo{author}{Rabani, E.}, \bibinfo{author}{Hetenyi, B.},
  \bibinfo{author}{Berne, B.~J.} \& \bibinfo{author}{Brus, L.~E.}
\newblock \bibinfo{title}{{Electronic properties of CdSe nanocrystals in the
  absence and presence of a dielectric medium}}.
\newblock \emph{\bibinfo{journal}{J. Chem. Phys.}}
  \textbf{\bibinfo{volume}{110}}, \bibinfo{pages}{5355--5369}
  (\bibinfo{year}{1999}).

\bibitem{Williamson2000}
\bibinfo{author}{Williamson, A.} \& \bibinfo{author}{Zunger, A.}
\newblock \bibinfo{title}{{Pseudopotential study of electron-hole excitations
  in colloidal free-standing InAs quantum dots}}.
\newblock \emph{\bibinfo{journal}{Phys. Rev. B}} \textbf{\bibinfo{volume}{61}},
  \bibinfo{pages}{1978--1991} (\bibinfo{year}{2000}).

\bibitem{Wall1995}
\bibinfo{author}{Wall, M.~R.} \& \bibinfo{author}{Neuhauser, D.}
\newblock \bibinfo{title}{{Extraction, through filter-diagonalization, of
  general quantum eigenvalues or classical normal mode frequencies from a small
  number of residues or a short-time segment of a signal. I. Theory and
  application to a quantum-dynamics model}}.
\newblock \emph{\bibinfo{journal}{J. Chem. Phys.}}
  \textbf{\bibinfo{volume}{102}}, \bibinfo{pages}{8011--8022}
  (\bibinfo{year}{1995}).

\bibitem{Toledo2002}
\bibinfo{author}{Toledo, S.} \& \bibinfo{author}{Rabani, E.}
\newblock \bibinfo{title}{{Verly large electronic structure calculations using
  an out-of-core filter-diagonalization method}}.
\newblock \emph{\bibinfo{journal}{J. Comput. Phys.}}
  \textbf{\bibinfo{volume}{180}}, \bibinfo{pages}{256--269}
  (\bibinfo{year}{2002}).

\bibitem{Shabaev2004}
\bibinfo{author}{Shabaev, A.} \& \bibinfo{author}{Efros, A.~L.}
\newblock \bibinfo{title}{{1D exciton spectroscopy of semiconductor nanorods}}.
\newblock \emph{\bibinfo{journal}{Nano Lett.}} \textbf{\bibinfo{volume}{4}},
  \bibinfo{pages}{1821--1825} (\bibinfo{year}{2004}).

\bibitem{Rabani2010}
\bibinfo{author}{Rabani, E.} \& \bibinfo{author}{Baer, R.}
\newblock \bibinfo{title}{{Theory of multiexciton generation in semiconductor
  nanocrystals}}.
\newblock \emph{\bibinfo{journal}{Chem. Phys. Lett.}}
  \textbf{\bibinfo{volume}{496}}, \bibinfo{pages}{227--235}
  (\bibinfo{year}{2010}).

\bibitem{Baer2012}
\bibinfo{author}{Baer, R.} \& \bibinfo{author}{Rabani, E.}
\newblock \bibinfo{title}{{Expeditious stochastic calculation of multiexciton
  generation rates in semiconductor nanocrystals}}.
\newblock \emph{\bibinfo{journal}{Nano Lett.}} \textbf{\bibinfo{volume}{12}},
  \bibinfo{pages}{2123--2128} (\bibinfo{year}{2012}).

\bibitem{Cunningham2011}
\bibinfo{author}{Cunningham, P.~D.} \emph{et~al.}
\newblock \bibinfo{title}{{Enhanced multiple exciton generation in
  quasi-one-dimensional semiconductors}}.
\newblock \emph{\bibinfo{journal}{Nano Lett.}} \textbf{\bibinfo{volume}{11}},
  \bibinfo{pages}{3476--3481} (\bibinfo{year}{2011}).

\bibitem{Sandberg2012}
\bibinfo{author}{Sandberg, R.~L.} \emph{et~al.}
\newblock \bibinfo{title}{{Multiexciton dynamics in infrared-emitting colloidal
  nanostructures probed by a superconducting nanowire single-photon detector}}.
\newblock \emph{\bibinfo{journal}{ACS Nano}} \textbf{\bibinfo{volume}{6}},
  \bibinfo{pages}{9532--9540} (\bibinfo{year}{2012}).

\bibitem{Baer2013}
\bibinfo{author}{Baer, R.} \& \bibinfo{author}{Rabani, E.}
\newblock \bibinfo{title}{{Communication: Biexciton generation rates in CdSe
  nanorods are length independent}}.
\newblock \emph{\bibinfo{journal}{J. Chem. Phys.}}
  \textbf{\bibinfo{volume}{138}}, \bibinfo{pages}{051102}
  (\bibinfo{year}{2013}).

\bibitem{Barzykin2005}
\bibinfo{author}{Barzykin, A.~V.} \& \bibinfo{author}{Tachiya, M.}
\newblock \bibinfo{title}{{Stochastic models of carrier dynamics in
  single-walled carbon nanotubes}}.
\newblock \emph{\bibinfo{journal}{Phys. Rev. B}} \textbf{\bibinfo{volume}{72}},
  \bibinfo{pages}{075425} (\bibinfo{year}{2005}).

\bibitem{Ma2005}
\bibinfo{author}{Ma, Y.~Z.}, \bibinfo{author}{Valkunas, L.},
  \bibinfo{author}{Dexheimer, S.~L.}, \bibinfo{author}{Bachilo, S.~M.} \&
  \bibinfo{author}{Fleming, G.~R.}
\newblock \bibinfo{title}{{Femtosecond spectroscopy of optical excitations in
  single-walled carbon nanotubes: Evidence for exciton-exciton annihilation}}.
\newblock \emph{\bibinfo{journal}{Phys. Rev. Lett.}}
  \textbf{\bibinfo{volume}{94}}, \bibinfo{pages}{157402}
  (\bibinfo{year}{2005}).

\bibitem{Huang2006}
\bibinfo{author}{Huang, L.} \& \bibinfo{author}{Krauss, T.~D.}
\newblock \bibinfo{title}{{Quantized bimolecular auger recombination of
  excitons in single-walled carbon nanotubes}}.
\newblock \emph{\bibinfo{journal}{Phys. Rev. Lett.}}
  \textbf{\bibinfo{volume}{96}}, \bibinfo{pages}{057407}
  (\bibinfo{year}{2006}).

\bibitem{Wang2011}
\bibinfo{author}{Wang, F.}, \bibinfo{author}{Dukovic, G.},
  \bibinfo{author}{Brus, L.~E.} \& \bibinfo{author}{Heinz, T.~F.}
\newblock \bibinfo{title}{{The Optical Resonances in Carbon}}.
\newblock \emph{\bibinfo{journal}{Science}} \textbf{\bibinfo{volume}{308}},
  \bibinfo{pages}{838--841} (\bibinfo{year}{2005}).

\bibitem{Pal2017}
\bibinfo{author}{Pal, S.}, \bibinfo{author}{Casanova, D.} \&
  \bibinfo{author}{Prezhdo, O.~V.}
\newblock \bibinfo{title}{{Effect of Aspect Ratio on Multiparticle Auger
  Recombination in Single-Walled Carbon Nanotubes: Time Domain Atomistic
  Simulation}}.
\newblock \emph{\bibinfo{journal}{Nano Lett.}} \textbf{\bibinfo{volume}{18}},
  \bibinfo{pages}{58--63} (\bibinfo{year}{2018}).

\end{thebibliography}

\begin{addendum}
\item [Supplementary Information] is available for this paper at https://doi.org/.
\item [Acknowledgements] The authors thank Mr. Devan Skubitz for preliminary
tests of the developed code and Dr. Felipe H. Jornada and Prof. Steven
G. Louie for stimulating discussions. This research was supported
by the University of California Lab Fee Research Program (Grant LFR-17-477237)
and used resources of the National Energy Research Scientific Computing
Center (NERSC), a U.S. Department of Energy Office of Science User
Facility operated under Contract No. DE-AC02-05CH11231.
\item [Author Contributions] J.P.P and E.R. developed the theoretical framework,
computer code, performed the calculations and co-wrote the paper.
\item [Competing Interests] The authors declare that they have no competing
financial interests. 
\item [Correspondence] Correspondence and requests for materials should
be addressed to J.P.P.~(email: jphilbin@berkeley.edu) or to E.R.~(email:
eran.rabani@berkeley.edu).
\end{addendum}

\end{document}